\documentclass[%
 aip,
 amsmath,amssymb,
twocolumn,%
twoside,%
 reprint,%
]{revtex4-1}

\usepackage{graphicx}

\begin{document}


\title{Quantized optical near-field interactions measured with a superconducting nanowire detector} 



\author{Karol Luszcz}
\affiliation{ETH Z\"urich, Photonics Laboratory, 8093 Z\"urich, Switzerland}
\author{Eric Bonvin}
\affiliation{ETH Z\"urich, Photonics Laboratory, 8093 Z\"urich, Switzerland}
\author{Lukas Novotny}
\affiliation{ETH Z\"urich, Photonics Laboratory, 8093 Z\"urich, Switzerland}


\date{\today}

\begin{abstract}
The quantum nature of optical near-fields is interesting from a theoretical perspective and of importance for practical applications, such as high-resolution imaging, sensing and antenna-coupled quantum light sources.
In this work we use a custom-designed superconducting nanowire single-photon detector (SNSPD) to directly read out the near-field interaction between source and detector. We use a subwavelength-sized aperture at the end of an optical fiber to record spatial near-field maps and to measure the distance dependence of the optical near-field interaction. Our measurements can be well described by a superposition of evanescent source fields with no noticeable probe-sample coupling. Our approach can be used for further studies of quantum near-field interactions and for the development of near-field imaging techniques with single quantum sensitivity.  
\end{abstract}


\maketitle 

\section{Motivation}
The quantum nature of optical near-fields has been of interest for years~\cite{carniglia71,vigoureux80}. More recent studies focused on the coupling of optical antennas to single-photon detecting devices~\cite{heath15,hu11,csete13,wang17} and the characterization of single-photon detection in nanoscale environments~\cite{engel15b,renema15}.\\

A single photon detector records the arrival of photons, quanta of propagating electromagnetic radiation. Propagating radiation is transverse and it is decoupled from the source that generated it. On the other hand, optical near-fields are intimately coupled to their sources. Consequently, when a single photon detector interacts with an optical near-field the detector's clicks represent quanta of the electromagnetic interaction between source and detector and \textit{not} the electromagnetic field per se.~\cite{carniglia71, power97, keller12} Thus, in this context, it is of interest to investigate and understand the interaction of optical near-fields with single photon detectors.\\

For the investigation of optical near-fields in the quantum regime, a method for direct probing of the field-matter interaction in the near-field is required. Superconducting Nanowire Single-Photon Detectors (SNSPDs)~\cite{goltsman01,natarajan12} can be fabricated in the form of a subwavelength-sized area with single-photon sensitivity,~\cite{bitauld10} which qualifies them for \textit{direct} probing of near-field interactions, unlike traditional scanning near-field optical microscopy (SNOM) techniques. In SNOM,  the near-field is typically converted by a local probe into propagating radiation, which is then guided to a single-photon detector, either by an optical fiber or through free space. Thus, the detector interacts with optical far-fields and the optical near-field interaction is \textit{not} directly measured.\\

To directly detect the optical near-field interaction we here place a single-photon detector close to a subwavelength aperture, at a distance of a few nanometers. No propagating radiation is involved in the detection process. In addition to characterizing the properties of the near-field interaction, the ability to directly probe evanescent fields, without the need to scatter the evanescent field into propagating radiation, is a promising approach for improving the sensitivity of SNOM and for expanding the applications of this technique to new areas. The presented work is a step towards developing a nanoscale single-photon detector integrated onto a scanning tip \cite{wang13,wang15,hayden06}.

\begin{figure}[!b]
	\includegraphics[width=0.48\columnwidth]{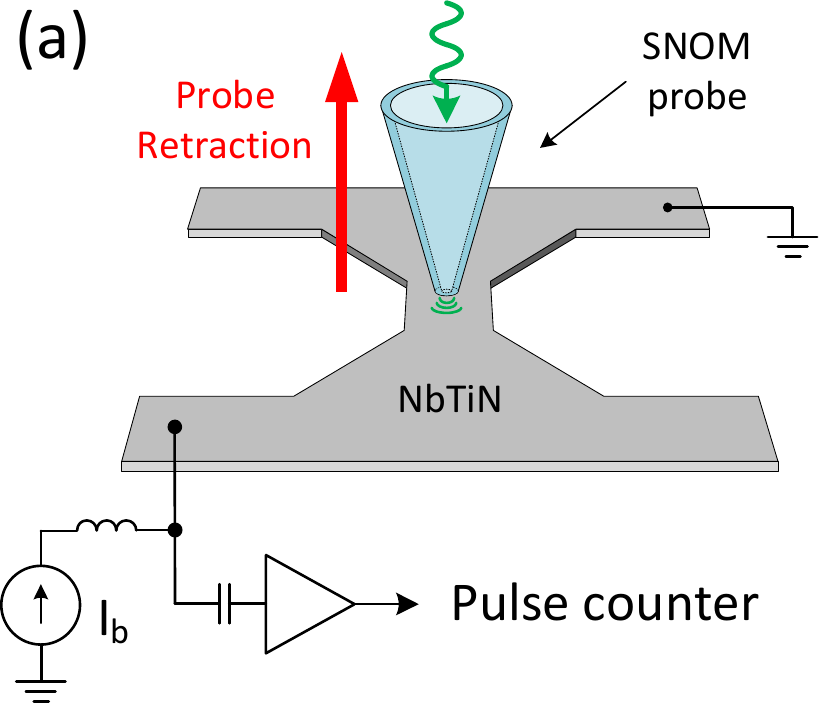}%
	~
	\includegraphics[width=0.51\columnwidth]{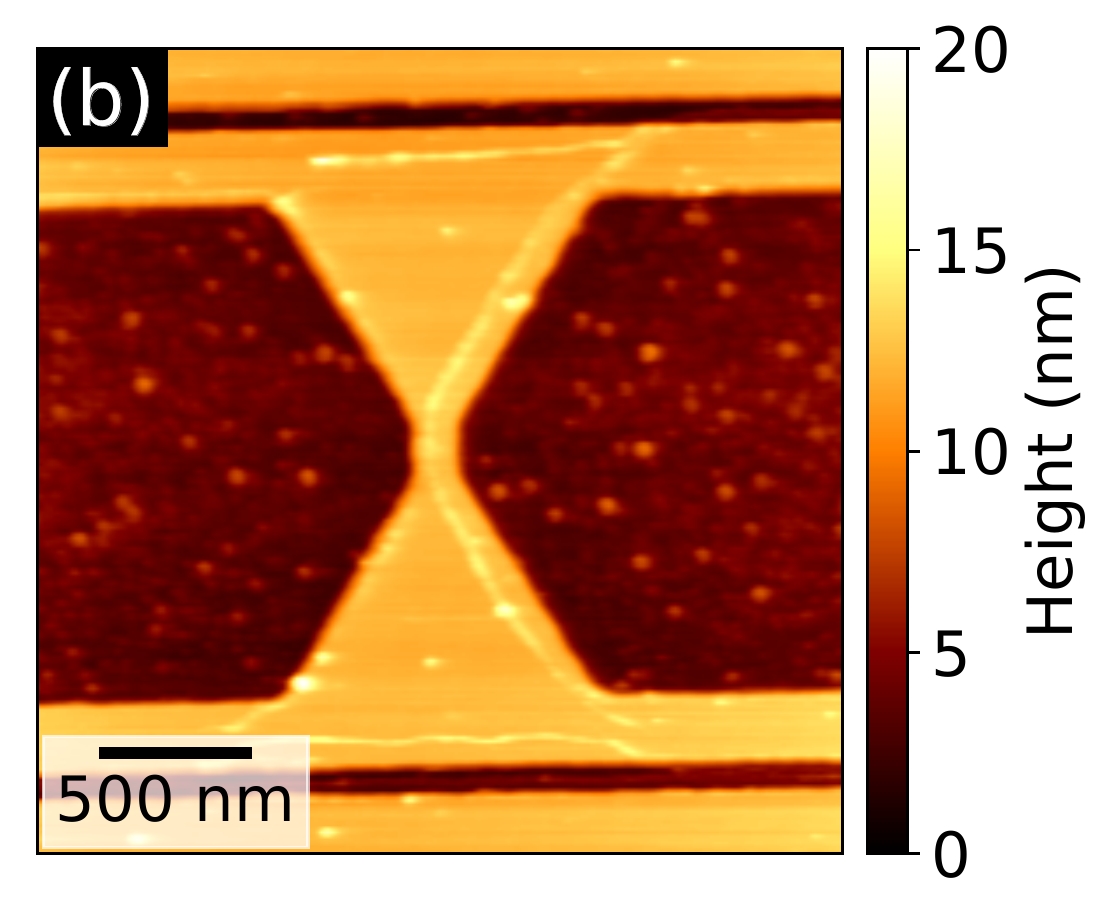}%
	\caption{(a) Illustration of the experiment. A subwavelength aperture acts as the source of the optical near-field. The aperture is placed in close proximity over an SNSPD. Detector clicks are recorded as a function of the aperture position. (b) Atomic force microscopy (AFM) image of the SNSPD. The device features a 200\,$\times$\,200\,nm$^2$ constriction in a 6\,nm-thick NbTiN film deposited on sapphire substrate. The pattern was fabricated by electron beam lithography followed by reactive ion etching.\label{fig:Experiment_sketch}}
\end{figure}

\begin{figure}[!b]
	\includegraphics[width=0.8\columnwidth]{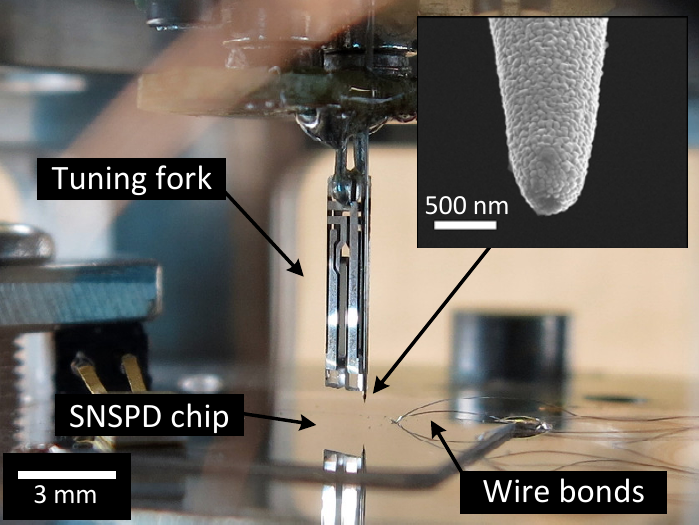}
	\caption{A close-up photograph of the experimental configuration. The SNOM aperture probe is attached to a quartz tuning fork and positioned above the SNSPD chip. The inset shows a scanning electron microscopy image of a SNOM aperture probe.\label{fig:tip_above_snspd}}
\end{figure}
\section{Experimental Configuration}
A sketch of the experiment is shown in Fig.~\ref{fig:Experiment_sketch}a. The SNSPD  is operated at a bias current of approximately\ 7\,$\mathrm{\mu}$A.  The electrical pulses generated by the near-field interaction are amplified by two \mbox{wide-band} RF amplifiers connected in series and converted to TTL digital standard with an adjustable threshold discriminator. The detector used in this work, shown in Fig.~\ref{fig:Experiment_sketch}b, is a 200\,$\times$\,200\,nm$^2$-area constriction in a \mbox{6\,nm-thick} NbTiN film. The device is fabricated on a sapphire substrate by magnetron sputtering thin-film deposition followed by electron beam lithography and reactive ion etching. \\

The detector is installed on a custom-built piezoelectric positioning stage. The stage together with the detector and the SNOM aperture probe is placed inside a  bath cryostat and submerged in superfluid helium at a temperature of approximately\ 2\,K. Once cold, the aperture probe is positioned directly on top of the SNSPD nanowire constriction. The aperture SNOM probes are fabricated by chemical etching of a glass fiber and subsequent Al evaporation. The probe-sample distance is sensed with a quartz tuning-fork based FM shear-force feedback mechanism~\cite{karrai95,atia97} and maintained constant with a piezoelectric tube actuator. The tracking of the tuning fork resonance frequency is realized with a commercial phased-locked loop. The control of the probe (positioning and scanning) is handled by a commercial SPM controller. \\

Fig.~\ref{fig:tip_above_snspd} shows a photograph of the SNSPD detector chip underneath the SNOM probe attached to a quartz tuning fork. In the research presented in this paper, two different types of aperture probes and wavelengths were used. For distance-dependent near-field measurements, we a wavelength of 532\,nm and for near-field imaging we used a wavelength of 700\,nm. The power at the input of the SNOM fibers was approx.\ 0.5\,mW in both cases.

\section{Near-field imaging}
\begin{figure}[!t]
	\includegraphics[height=1.42in]{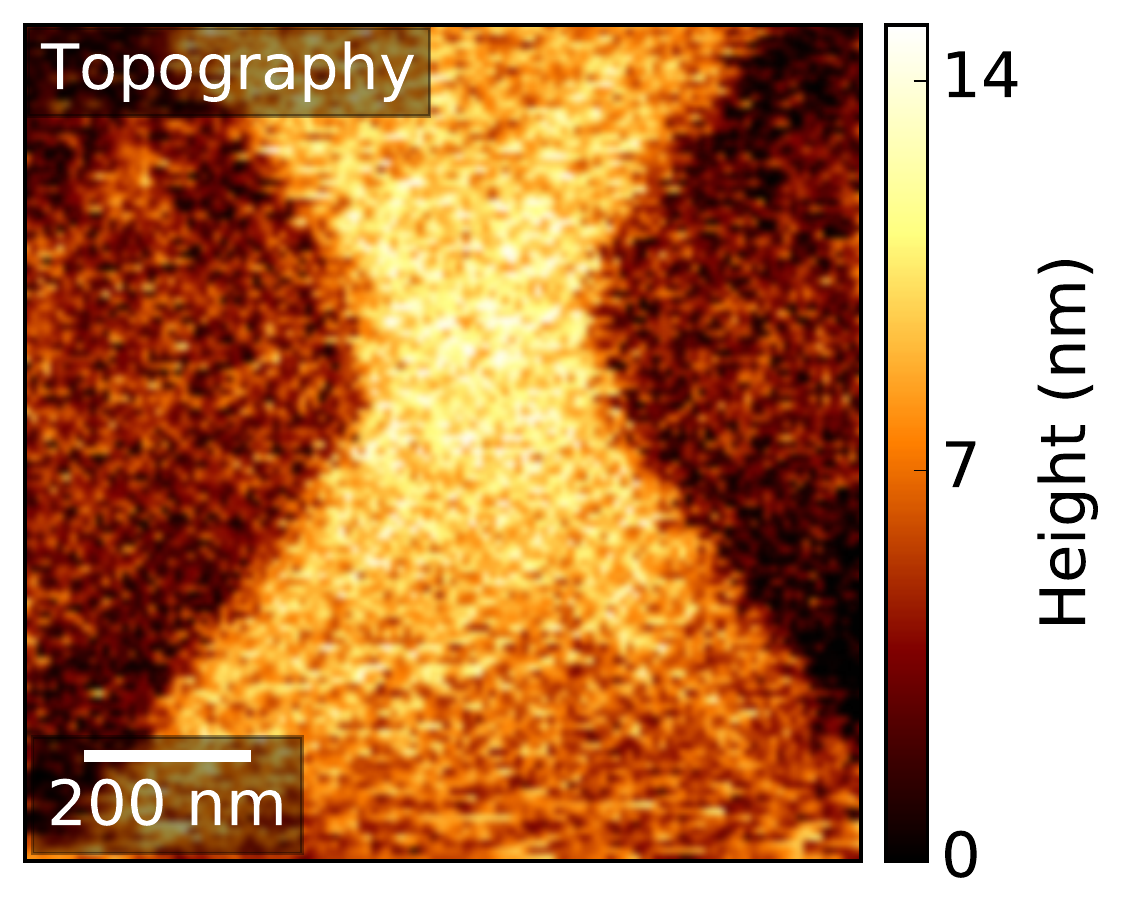}%
	~~~~
	\includegraphics[height=1.42in]{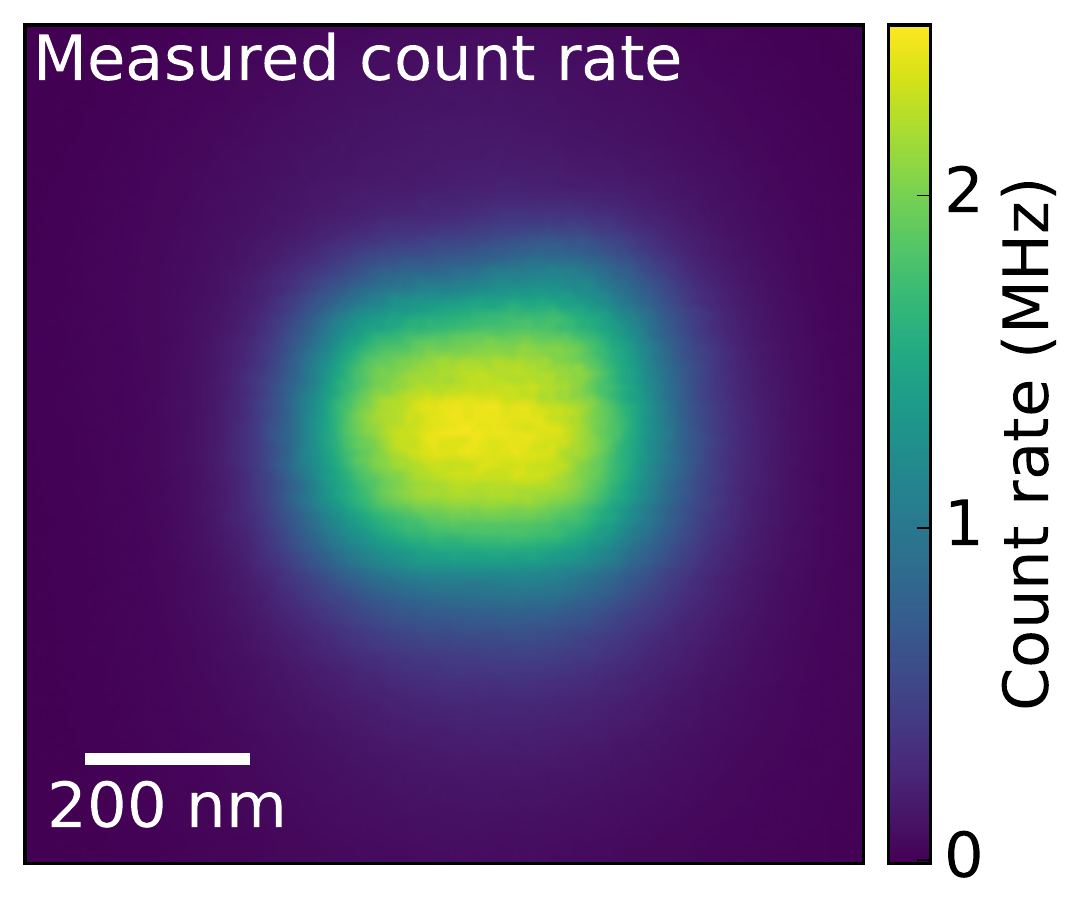}%
	\caption{Topography (left) and count rate (right) images of the SNSPD constriction area acquired by raster-scanning the SNOM probe over the surface of the SNSPD. The two images were acquired simultaneously in superfluid helium at a temperature of 2\,K using a SNOM probe with a 220\,nm-size aperture.\label{fig:Map}}%
\end{figure}

After cooling the experimental setup and submerging it in superfluid helium, we approach the aperture probe to the SNSPD and record raster-scan images while maintaining a constant probe-sample separation of a few nanometers. While scanning, we record the rate of pulses generated by the SNSPD detector. This allows us to record the spatial sensitivity distribution of the detector surface with subdiffraction resolution. Simultaneously we record the corresponding topographical map. Fig.~\ref{fig:Map} shows 1\,$\times$\,1\,$\mathrm{\mu}$m$^2$ images of the SNSPD constriction zone for both the count rate and topography. The two images were recorded in superfluid helium with a 220\,nm-size aperture probe. The acquired images confirm that the SNSPD detector is light-sensitive only in the nanowire constriction area as it is expected, and that the SNOM probe can be accurately positioned in the center of that area. The positioning accuracy needed for the distance-dependent measurements is discussed in the following section.

\section{Probe-sample distance dependence}
\begin{figure}[!t]
\includegraphics[width=\columnwidth]{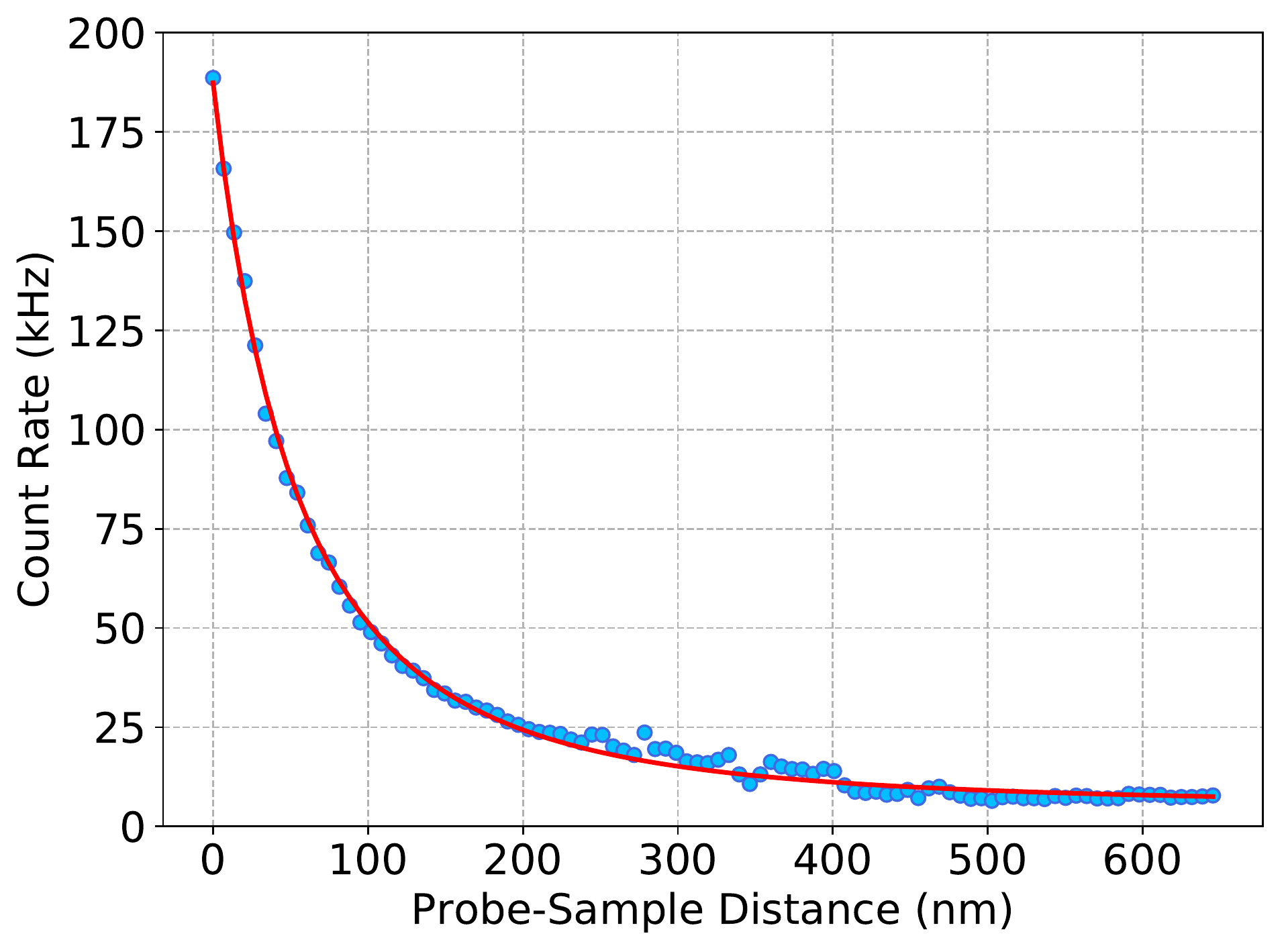}%
\caption{Dependence of the optical near-field interaction on probe-sample distance. Dots are measurements and solid lines are theoretical fits. The aperture probe was retracted from the sample at a speed of 50\,nm/s. \label{fig:appr_curve_532}}%
\end{figure}

The count rate of the pulses recorded at the output of the SNSPD detector represents the strength of the quantum interaction between the source (aperture probe) and the detector (SNSPD nanowire). By positioning the SNOM probe in the center of the nanowire constriction and retracting it away from the surface of the sample we record the distance dependence of the near-field. As shown in Fig.~\ref{fig:appr_curve_532} we observe a nearly-exponential decay, characteristic of an evanescent wave. \\

To theoretically describe the measured behavior we model the source by a uniform in-plane electric field $E_0$  constrained to a circular area of radius $w_0$ located in the plane $z$\,=\,0. Following Novotny and Hecht~\cite{novotny12}, the field spreads out as
\begin{equation}
E_x(x,y,z) =  E_0 \frac{w_0^2}{2} \int_0^\infty e^{-\frac{1}{4} k_\parallel^2 w_0^2} k_{\parallel} J_0(k_{\parallel}\rho) e^{\mathrm{i} k_z z} dk_{\parallel},
\label{eq:Integral}
\end{equation}
where $J_0$ is the Bessel function of the first kind, \mbox{$k_{\parallel}=\sqrt{k_x^2+k_y^2}$} is the \mbox{in-plane} spatial frequency, and $k_z$ is the spatial frequency perpendicular to the aperture plane:
\small\begin{equation}
k_z = \sqrt{\left(\frac{2 \pi}{\lambda}\right)^2 - k_{\parallel}^2}.
\label{eq:Ch3-kz}
\end{equation}\normalsize
For an aperture of a size much smaller than the wavelength of illuminating radiation $k_z$ is mainly imaginary. Therefore, the field behind the aperture is primarily composed of non-propagating near-fields that decay exponentially from the aperture plane.\\

In Fig.~\ref{fig:appr_curve_532}, we superimpose the intensity of the field (\ref{eq:Integral}) to the measured detector count rate. Excellent agreement between theory and experiment is obtained. The fitting parameters are the aperture size $w_0$ and the field intensity in the aperture plane $E_0$ (related to the maximum count rate at shortest probe-sample distance). The field decays exponentially and the aperture radius resulting from the fitting process is $w_0$\,=\,94\,nm. The aperture radius is in agreement with the probe specifications and corresponds to $\lambda$/5.66 for a wavelength of 532\,nm.\\

The results shown in Fig.~\ref{fig:appr_curve_532} seem to indicate that the SNSPD clicks represent the classical electromagnetic field of the source and that the SNSPD acts as a passive probe. We observe no measurable backaction from the detector on the source, which would result in interference undulations, akin to Fabry-P\'erot resonances with a period of $\lambda$/2\,=\,266\,nm. We verified that such interference undulations do exist for near-field probes with larger aperture size and hence larger contribution of far-field radiation. Furthermore, evanescent coupling between probe and sample should have its signature in a flattening of the distance dependence at small distances, as is the case  in frustrated total internal reflection~\cite{novotny12}. No such behavior is observed in our data shown Fig.~\ref{fig:appr_curve_532}.  Thus, our data can be best described by a superposition of exponentially decaying near-fields and the contribution of probe-sample coupling can be largely ignored.

\section{Conclusions}
In this work we measured the quanta of the optical near-field interaction between a source and a detector. We used a custom, subwavelength-size SNSPD detector directly interacting with a subwavelength aperture of a SNOM probe. We record spatial near-field maps and measure the distance dependence of the near-field interaction at a temperature of 2\,K. We find that the detector signal can be well described by a superposition of evanescent source fields and that the backaction on the source can be neglected. The experimental platform developed in this research can be used for further investigation of the quantum nature of optical near-field interactions and is a step towards the development of novel SNOM techniques for ultrasensitive super-resolution imaging, in particular, integration of a subwavelength single-photon detector on a scanning tip.

\begin{acknowledgments}
We thank Andreas Schilling, Andreas Engel, and Martin Frimmer for valuable support and fruitful discussions. This work has been supported by ETH Grant ETH-07 16-1.\\
\end{acknowledgments}


%

\end{document}